\documentclass[preprint,12pt]{elsarticle}

\usepackage{amssymb}
\usepackage{graphicx}
\usepackage{float}

\begin{document}
\begin{frontmatter}

\author[INT]{Maciej J. Winiarski}
\author[UWr]{Kaja Bili\'nska}
\author[INT]{Kamil Ciesielski}
\author[INT]{Dariusz Kaczorowski}
\address[INT]{Institute of Low Temperature and Structure Research, Polish Academy of Sciences, Ok\'olna 2, 50-422 Wroc\l aw, Poland}
\address[UWr]{Institute of Theoretical Physics, University of Wroc\l aw, Plac M. Borna 9, 50-204 Wroc\l aw, Poland}

\title{Thermoelectric performance of $p$-type half-Heusler alloys Sc$M$Sb ($M$ = Ni, Pd, Pt) by {\it ab initio} calculations}

\begin{abstract}
Structural, electronic, and transport properties of ScNiSb, ScPdSb, and ScPtSb were investigated from first principles. Electronic band structures derived within the fully relativistic MBJLDA approach were compared with those obtained from the standard GGA calculations. All the compounds studied exhibit indirect narrow band gaps (0.24--0.63~eV). The effective masses of hole-like carriers are relatively small (0.27--0.36),  and decrease with an increasing atomic number of the transition metal component. The carrier relaxation time, required for realistic calculations of the electrical conductivity, was approximated within the deformation potential theory. The GGA approach yielded overestimated transport characteristics with respect to those derived within the MBJLDA analysis. The largest power factor of 4--6~mWK$^{-2}$m$^{-1}$) at high temperatures was obtained for ScPtSb. This value is comparable with those observed experimentally for Fe-Nb-Sb half-Heusler alloys, and hence makes ScPtSb a very good candidate 
material for thermoelectric applications.
\end{abstract}


\end{frontmatter}

\section{Introduction}

Thermoelectric properties of Sb-based half-Heusler alloys have drawn attention to possible applications of these systems, e.g., for waste-heat recovery in heavy industry. Recently, a high figure of merit in heavy-band $p$-type Fe-Nb-Sb materials doped with some transition metals was extensively investigated \cite{Fu, Bauer, Cava, Joshi, Fu2, He}. In turn, available experimental data for another series of half-Heusler antimonides Sc-$M$-Sb \cite{Oestreich,Harmening}, where $M$ is Ni, Pd, or Pt, is very limited. In a single experimental study on thermoelectric performance of these materials \cite{Oestreich}, fairly small values of the figure of merit $ZT$ at 300 K were reported, and the authors suggested the presence of very narrow band gaps in their electronic band structures.

Electrical transport characteristics can be theoretically investigated in a semi-classical {\it ab initio} manner using the BolzTraP code \cite{Boltztrap}. Furthermore, calculations based on the density functional theory (DFT) may provide a valuable guidance to experimental work on the thermoelectric properties of half-Heusler alloys \cite{DFTAdv}. Particularly, the Seebeck coefficients and the power factors of Y-$M$-Sb systems ($M$ = Ni, Pd, Pt) were investigated within such an approach \cite{YMSb,YPdSb}. However, stand-alone band structure calculations are insufficient to obtain the relaxation time of charge carriers, and hence conclusions derived within the simple DFT approach are merely qualitative. To account for this deficiency, in some recent works on half-Heusler materials \cite{DFT0,DFT1,DFT2,DFT3,DFT4,DFT5}, the deformation potential theory \cite{Bardeen} was employed to get an estimate of the relaxation time at high temperatures, where scattering of charge carriers is governed by acoustic phonons. 
This way, the thermoelectric performance of these compounds was predicted in a quantitative manner. However, most recently, Guo \cite{Guo} showed that the common calculation scheme within the scalar relativistic approach, used for investigations of such heavy-band systems, leads to significant overestimation of the power factors in $p$-type regime. This effect is related to different shapes of the heavy- and light-hole bands obtained with the fully relativistic and scalar relativistic calculations.

In this work, the fully relativistic DFT-based approach and the carrier relaxation time approximation within the deformation potential theory were applied to calculate the electronic band structures and the thermoelectric properties of three $p$-type half-Heusler compounds ScNiSb, ScPdSb and ScPtSb. Because of the well-known underestimation of band gap values by standard parameterizations of the exchange-correlation energy, the modified Becke-Johnson potential (MBJLDA \cite{MBJLDA}) was used for examination of the reference GGA-derived \cite{GGA} results. For each material, the thermoelectric power factor (PF) was evaluated as a function of the carrier concentration and its temperature dependence was derived. The results obtained should encourage experimental research on the high-temperature thermoelectric properties of these phases.

\section{Computational details}

The electronic structure calculations were performed using the full-potential Wien2k package \cite{Wien2k}. The  Perdew-Burke-Erzerhof (GGA) \cite{GGA} and MBJLDA (TB-mBJ) \cite{MBJLDA} parameterizations of the exchange-correlation energy were employed. $RK_{max}$ = 8 and a $12\times12\times12$ {\bf k}-point mesh were used. The spin-orbit coupling was included. The thermoelectric characteristics were calculated with a 200000 {\bf k}-point mesh using the BolzTraP code \cite{Boltztrap}. Following the deformation potential theory \cite{Bardeen}, the relaxation time along $\beta$ direction at temperature $T$ was expressed as

$$ \tau_\beta = \frac{2 \sqrt{2\pi} C_\beta \hbar^4}{ 3(k_B Tm_{DOS}*)^{3/2} E_\beta^2 }, $$

where $C_{\beta}$ stands for the corresponding elastic constant, $m_{DOS}^*=(m_x^* m_y^* m_z^*)^{1/3}$ represents the density of states (DOS) effective mass, derived for heavy- and light-hole bands, and $E_{\beta}=\frac{\partial E_{VBM}}{\partial(\Delta l / l_0)}$ is the deformation potential that is a change of the valence band maximum (VBM) energy $E_{VBM}$ upon strain application ($l_0$ and $\Delta l$ denote the diffusion length and its relative change, respectively).

\section{Results and discussion}

The equilibrium structural parameters of the Sc$M$Sb compounds investigated are gathered in Table \ref{table1}. It should be noted that the lattice parameters obtained within the GGA approach are overestimated with respect to the experimental data by 1.2--1.5~\%. Interestingly, the calculated bulk modulus $B$ does not follow an increasing atomic number of the $M$ component on going from Ni to Pd to Pt. The smallest value of $B$ was found for ScPdSb, whereas the largest one was derived for ScPtSb.

\begin{table}
\caption{Cubic lattice parameter $a$ and bulk modulus $B$ derived using GGA for ScNiSb, ScPdSb and ScPtSb. The parameter $\Delta a$ is the relative difference between the calculated value of $a$ and the experimental value reported in Ref. \cite{Oestreich}.}
\label{table1}
\begin{tabular}{llll}
\hline
Compound & $a$ (A) & $\Delta a$ (\%) & $B$ (GPa) \\ \hline
ScNiSb & 6.138 & 1.25 & 106.03  \\
ScPdSb & 6.392 & 1.27 & 97.42  \\
ScPtSb & 6.405 & 1.48 & 113.72  \\
\end{tabular}
\end{table}

\begin{figure}
\includegraphics[width=8cm]{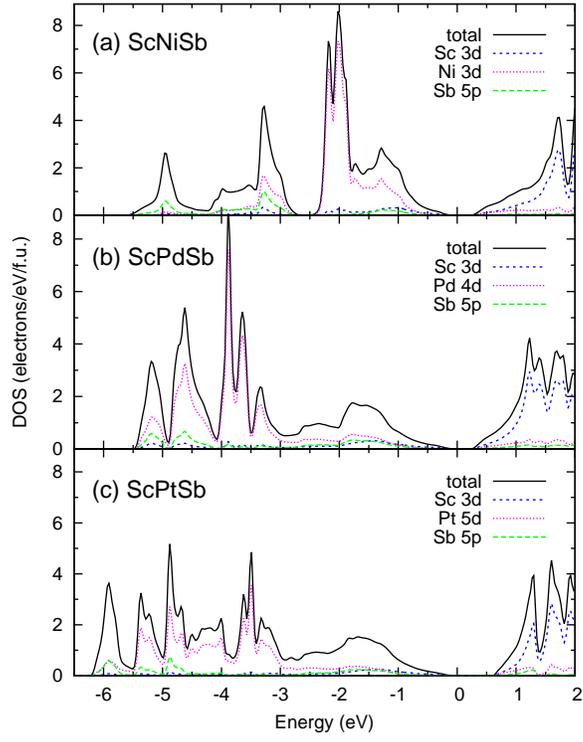}
\caption{Total density of states (DOS) and partial DOS contributions for (a) ScNiSb, (b) ScPdSb, and (c)  ScPtSb, calculated within the MBJLDA approach.}
\label{Fig1}
\end{figure}

The total density of states, calculated for the Sc$M$Sb systems using MBJLDA, is depicted in Fig. \ref{Fig1}. The shape of DOS reflects different chemical character of the transition metal atoms in these alloys. Namely, the main contributions of the Ni $3d$ states in ScNiSb are located in the energy region of 1--2 eV below VBM, whereas in the Pd- and Pt-based systems the corresponding $4d$/$5d$ bands are located at higher binding energies. As a result, in the latter two phases, total DOS in the vicinity of VBM is flattened in comparison to that in ScNiSb. In all three intermetallics, the conduction bands are dominated by the empty Sc $3d$ states and they are separated from the valence band by an energy gap $E_g$.

As can be inferred from Table \ref{table2}, the band gaps calculated for the Sc$M$Sb phases using GGA are fairly narrow. The values of $E_g^{GGA}$ obtained for ScNiSb and ScPdSb are similar, while $E_g^{GGA}$ found for ScPtSb is nearly twice larger. This finding gives rise to a presumption that the latter compound may exhibit high-temperature thermoelectric performance better that the other two materials, due to suppressed minority carrier activation over the semiconducting energy gap. It is worth recalling that the experimental values of $E_g$ determined for ScNiSb and ScPdSb are 0.11 and 0.23~eV, respectively \cite{Oestreich}. While the GGA calculations nicely reproduced the experimental estimate for the Pd-bearing compound, $E_g^{GGA}$ of ScNiSb is twice larger than the gap measured. The discrepancy may be attributed to structural disorder in the polycrystalline sample studied, which is a common feature for half-Heusler phases.

Table \ref{table2} comprises also the results of the MBJLDA calculations made for Sc$M$Sb. The value of $E_g^{MBJLDA}$ obtained for ScNiSb is fairly similar to the GGA result. For ScPdSb, $E_g^{MBJLDA}$ is somewhat bigger than $E_g^{GGA}$, while for ScPtSb it is significantly larger than the GGA value. Unlike the GGA data, MBJLDA yielded a systematic widening of the semiconducting gap on going from Ni to Pd to Pt, however the distinct jump in the value of $E_g^{MBJLDA}$ between ScPdSb and ScPtSb seems exaggerated.
Here, it is worth emphasizing that the electronic structure of half-Heusler alloys is generally exceptional considering that the presence of heavy atoms in common semiconductors, i.e., III-V materials \cite{Vurgaftman}, leads to flattening of valence bands and a decrease in $E_g$.

\begin{table}
\caption{Band gaps $E_g$ and heavy- and light- hole split-off energies $\Delta_{SO}$ (in eV) calculated for ScNiSb, ScPdSb and ScPtSb using GGA and MBJLDA.}
\label{table2}
\begin{tabular}{lllll}
\hline
Compound & $E_g^{GGA}$ & $\Delta_{SO}^{GGA}$ & $E_g^{MBJLDA}$ & $\Delta_{SO}^{MBJLDJ}$ \\ \hline
ScNiSb & 0.244 & 0.093 & 0.237 & 0.108 \\
ScPdSb & 0.229 & 0.189 & 0.261 & 0.181 \\
ScPtSb & 0.415 & 0.588 & 0.626 & 0.586 \\
\end{tabular}
\end{table}

\begin{figure*}
\includegraphics[width=15cm]{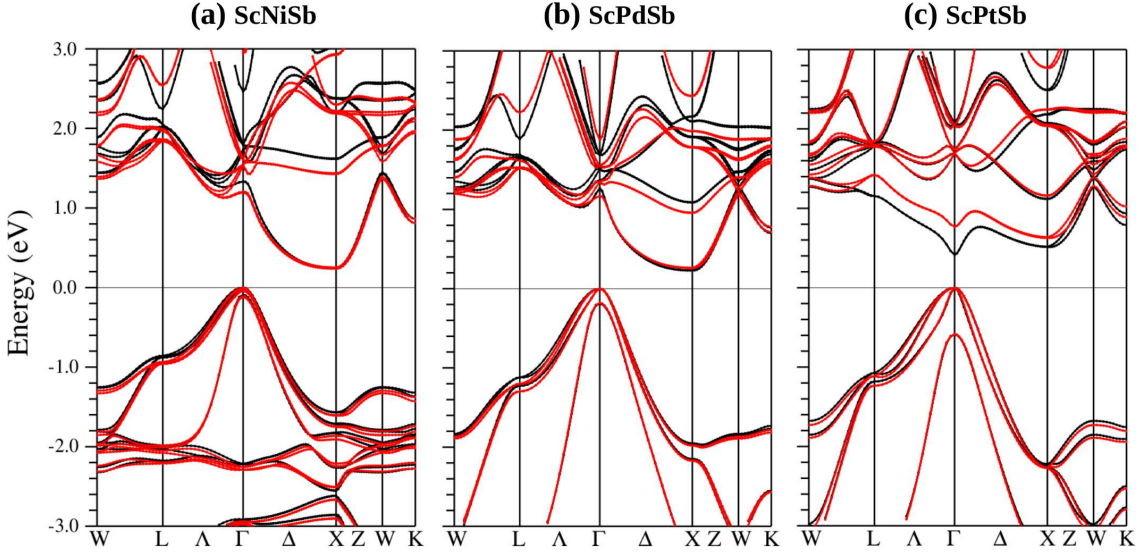}
\caption{Electronic band structures of (a) ScNiSb, (b) ScPdSb, and (c) ScPtSb, calculated within the GGA (black lines) and MBJLDA (red lines) approaches.}
\label{Fig2}
\end{figure*}

As shown in Fig. \ref{Fig2}, the GGA and MBJLDA calculations yielded similar shapes and positions of the valence bands in ScNiSb and ScPdSb. Also the positions of conduction band minimum (CBM) in the two compounds are almost the same from both methods. In contrast, CBM in ScPtSb derived by MBJLDA is strongly upshifted with respect to that found by GGA, and the energy gap $E_g^{MBJLDA}$ is of indirect type.
Interestingly, for the experimental value of the lattice parameter $a$ one obtained from MBJLDA a direct gap between the valence and conduction bands (not shown). This finding indicates that strain may cause a transition between direct and indirect band gap in ScPtSb. Analogous situation probably occurs for YPtSb that exhibits a band structure very similar to that of ScPtSb \cite{YMSb}, and its electrical transport behavior is strongly sample dependent \cite{YPtSb}. Regardless the differences in the shapes of CBM within the Sc$M$Sb series, rather high effective masses of the $n$-type carriers disqualify all three compounds as $n$-type thermoelectric materials \cite{Review}.

Regarding the valence bands in Sc$M$Sb, the split-off energies between heavy- and light-hole bands at the $\Gamma$ point of the Brillouin zone are proportional to the atomic number of transition metal atom $M$ (see Table \ref{table1}). It is worth noting that strong spin-orbit coupling in ScPtSb, reflected in the clearly pronounced splittings of the heavy-hole bands along the $L$--$\Gamma$ line in the Brillouin zone (see Fig. \ref{Fig2} (c)), may notably affect the results of electrical transport calculations \cite{Guo}. Namely, the degeneracy of bands forming VBM is lifted and the light-hole bands are downshifted, implying that the calculated $p$-type thermoelectric performance is significantly different in relation to that derived from the scalar relativistic calculations (not discussed here).

The approximation of the relaxation time of carriers requires data which can be derived from the DFT-based calculations, as gathered for the Sc$M$Sb compounds in Table \ref{table3}. Contrary to irregular change in the bulk modulus along the series (see Table \ref{table1}), the elastic constants $C_{\beta}$ are roughly proportional to the atomic number of $M$. The calculated deformation potentials are relatively high in relation to $E_{\beta}$ of 15--20~eV reported in the literature for other half-Heusler antimonides \cite{DFT0, DFT2, DFT4}. The values of $E_{\beta}^{MBJLDA}$ are generally higher than $E_{\beta}^{GGA}$.
The effective masses of the $p$-type carriers are similar to those reported for LaPtSb \cite{DFT0} and NbFeSb \cite{DFT2}. Though $m_{eff}^{GGA}$ and $m_{eff}^{MJBLDA}$ are very close to each other, the relaxation time values $\tau^{GGA}$ at 300~K are larger than $\tau^{MBJLDA}$, because of the difference in $E_{\beta}$ obtained within the two approaches. The values of $\tau$ are generally small, which is also a characteristic feature of NbFeSb \cite{DFT2} and TiIrSb \cite{DFT5}.

\begin{table*}
\caption{Elastic constants $C_{\beta}$ (GGA results), deformation potentials $E_{\beta}$, density of states effective masses $m_{eff}$, and relaxation times $\tau$ calculated for ScNiSb, ScPdSb and ScPtSb using GGA and MBJLDA methods.}
\label{table3}
\begin{tabular}{llllllll}
\hline
Compound & $C_{\beta}$ (GPa) & $E_{\beta}^{GGA}$ (eV) & $E_{\beta}^{MBJLDA}$ (eV) & $m_{eff}^{GGA}$ & $m_{eff}^{MBJLDA}$ & $\tau^{GGA}$ (fs) & $\tau^{MBJLDA}$ (fs) \\ \hline
ScNiSb & 103 & 31.7 & 37.2 & 0.36 & 0.36 & 16.3 & 12.1 \\
ScPdSb & 173 & 30.4 & 32.8 & 0.31 & 0.31 & 38.0 & 32.7 \\
ScPtSb & 348 & 36.1 & 37.3 & 0.26 & 0.27 & 71.4 & 62.5 \\
\end{tabular}
\end{table*}

\begin{figure}
\includegraphics[width=8cm]{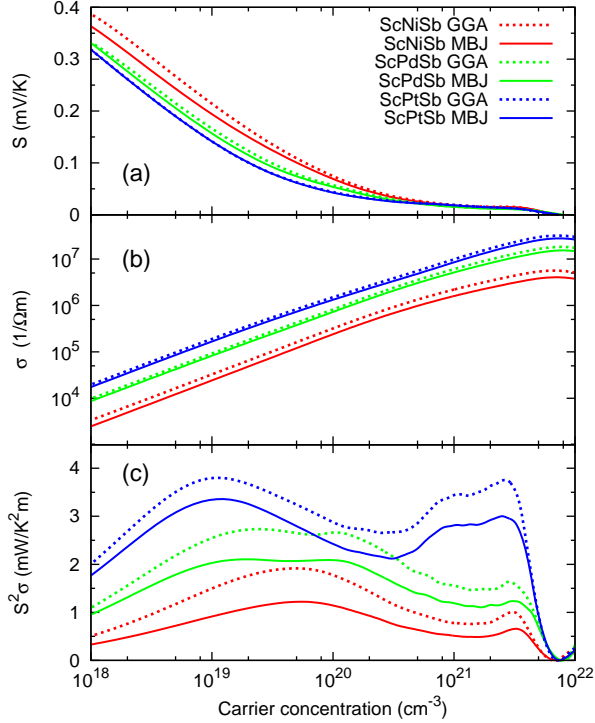}
\caption{(a) Seebeck coefficient $S$, (b) electrical conductivity $\sigma$, and (c) power factor $S^2\sigma$ at 300~K calculated for ScNiSb, ScPdSb and ScPtSb within the MBJLDA (solid lines) and GGA (dotted lines) approaches.}
\label{Fig3}
\end{figure}

The Seebeck coefficient $S$, electrical conductivity $\sigma$ and power factor PF = $S^2\sigma$, calculated for the Sc$M$Sb compounds as a function of the carrier concentration $n$ at 300~K, are presented in Figs. \ref{Fig3}(a), \ref{Fig3}(b) and \ref{Fig3}(c), respectively. In general, the $S(n)$ curves are similar for all three materials, showing a maximum value of $S$ = 0.2--0.3~mVK$^{-1}$ for small $n$ and then a systematic decrease towards zero with increasing $n$. At the same time, the electrical conductivity increases by about five orders of magnitude between the smallest and largest $n$ considered. In the entire range of the carrier concentration, the worst conductivity was found for ScNiSb, as a result of the shortest relaxation time $\tau$ (cf. Table \ref{table3}). In consequence, PF of this material is small, despite the largest values of $S$ within the Sc$M$Sb series. The power factor of ScPdSb is clearly higher than that of ScNiSb, whereas the superior PF values were predicted for ScPtSb.


According to the experimental studies on the Sc$M$Sb compounds \cite{Oestreich}, all these materials exhibit at room temperature very small values of $S$ (below 0.125~mVK$^{-1}$) and $\sigma$ ($10^4$--$10^6$ $\Omega^{-1}m^{-1}$). These data are in satisfactory quantitative agreement with the results calculated here for the regimes of medium and large $n$. Obviously, because the electronic structures of real samples are usually strongly affected by defects, grain boundaries, etc., an exact matching between the results of calculations made for ideal crystals and the experimental data cannot be expected.

\begin{figure}
\includegraphics[width=8cm]{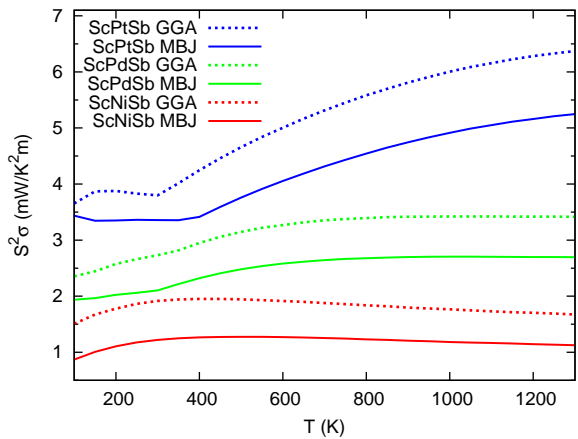}
\caption{Maximal power factor $S^2\sigma$ in a wide range of temperature, calculated for ScNiSb, ScPdSb and ScPtSb using the GGA (dotted lines) and MBJLDA (solid lines) approaches.}
\label{Fig4}
\end{figure}

Fig. \ref{Fig4} presents the calculated temperature variations of the maximal PF values of the Sc$M$Sb phases. For ScNiSb, this dependence is rather flat, whereas for the other two compounds, PF increases with increasing temperature. Remarkably, at 800~K, the values of PF derived for ScPtSb are comparable with those reported for the Fe-Nb-Sb materials \cite{Fu, Bauer, Cava, Joshi, Fu2, He}.

It is worth noting that the values of PF obtained within the MBJLDA approach are significantly smaller than those derived from the standard GGA calculations. This outcome is a direct consequence of the fact that the MBJLDA-derived magnitudes of $S$ and $\sigma$ are always smaller than the GGA data. Because there are no experimental reports on the electronic structure of the Sc$M$Sb compounds, one may assume that the MBJLDA and GGA results constitute the lower and upper reference limit, respectively, for the real materials. On one hand, the satisfactory agreement between the GGA-derived results and the experimental data for FeNbSb, quoted in Refs. \cite{DFT2,DFT4} for high-temperature regime, suggests that the standard DFT-based calculations may be an effective and suitable tool for predictions of the transport properties of some half-Heusler alloys. One the other hand, because the MBJLDA potential is generally considered superior for investigations of the band structures of semiconductors, one may expect 
that it is also most appropriate for calculations of the transport coefficients. Presently, however, this conjecture requires further experimental and theoretical research.

\begin{figure}
\includegraphics[width=8cm]{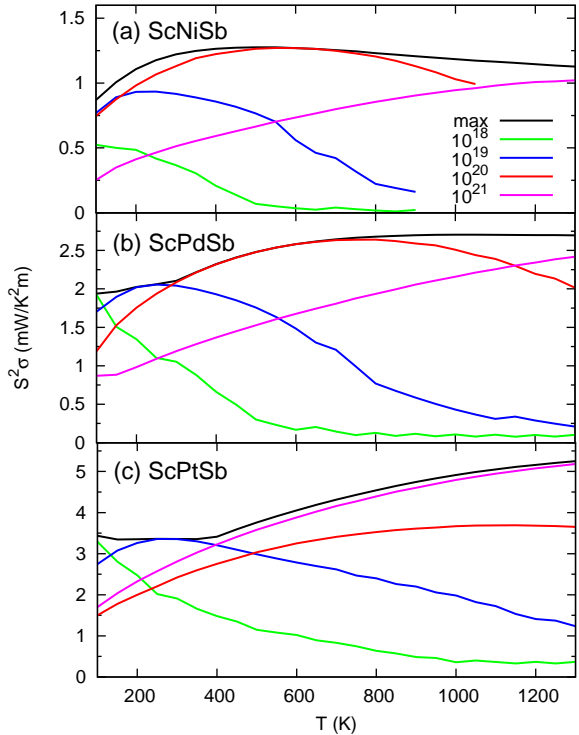}
\caption{Power factors $S^2\sigma$ for various carrier concentrations (in cm$^{-3}$) for (a) ScNiSb, (b) ScPdSb (b) and (c) ScPtSb, calculated within the MBJLDA approach. The maximal PF curve (labeled max) corresponds to optimal $n$, which generally varies with changing the temperature.}
\label{Fig5}
\end{figure}

Fig. \ref{Fig5} displays the temperature dependencies of the power factor of ScNiSb, ScPdSb and ScPtSb with different carrier concentrations. As can be inferred from this figure, for small values of $n$, PF always decreases with increasing temperature. The best thermoelectric performance of ScNiSb in a wide range of temperature is observed for $n$ of about 10$^{20}$~cm$^{-3}$. In turn, $n \approx$~10$^{19}$~cm$^{-3}$ provides optimal thermoelectric performance of ScPdSb and ScPtSb near room temperature, whereas larger values of $n$, namely 10$^{20}$ and 10$^{21}$~cm$^{-3}$, respectively, are desired at higher temperatures. Apparently, these two compounds are good candidate materials for thermoelectric applications, where their PF characteristics can be tuned by proper doping of $p$-type carriers.

\section{Conclusions}

The electronic structures of the half-Heusler alloys ScNiSb, ScPdSb and ScPtSb are fairly unique because of the very low effective masses of the $p$-type carriers. The latter compound is probably at the edge of a transition between the indirect and direct band gap, which probably could be tuned by application of strain. Most importantly, ScPtSb exhibits particularly good thermoelectric performance with the power factor reaching 4--6~mWK$^{-2}$m$^{-1}$ at high temperatures. For all the Sc$M$Sb phases studied, the thermoelectric characteristics can be tuned by intentional doping of charge carriers in a range from 10$^{19}$ to $10^{21} $~cm$^{-3}$.

The results obtained in this work indicate that using the deformation potential approximation for the carrier relaxation time is necessary for adequate discussion of the {\it ab initio} calculated transport properties of half-Heusler alloys. In view of the discrepancies between the data derived within the GGA and MBJLDA approaches, the most appropriate form of the exchange-correlation functional should be examined by comparing the theoretical predictions with the experimental data. For this purpose, experimental research on the thermoelectric properties of the Sc$M$Sb phases is presently underway.

\section*{Acknowledgments}
This work was partly supported by the National Science Centre (Poland) under research Grant no. 2015/18/A/ST3/00057.

\end{document}